\documentstyle[pre,aps]{revtex}

\makeatletter
\newfont{\bix}{cmbxti10}
\newfont{\bixi}{cmbxti10 scaled \magstephalf}
\newfont{\bixii}{cmbxti10 scaled \magstep1}
\newfont{\bixiv}{cmbxti10 scaled \magstep2}
\@addfontinfo\@xpt{\def\pbi{\bix}}
\@addfontinfo\@xipt{\def\pbi{\bixi}}
\@addfontinfo\@xiipt{\def\pbi{\bixii}}
\@addfontinfo\@xivpt{\def\pbi{\bixiv}}
\def\bi{\protect\pbi}
\makeatother

\begin{document}

\draft

\title{
  Pattern formation of reaction-diffusion system having
  self-determined flow in the amoeboid organism of
  {\bi Physarum} plasmodium
}

\author{
  Hiroyasu Yamada, Toshiyuki Nakagaki and Masami Ito
}
\address{
  Bio-Mimetic Control Research Centre,
  The Institute of Physical and Chemical Research (RIKEN)\\
  Shimoshidami, Moriyama, Nagoya, 463-0003, JAPAN
}



\maketitle


\begin{abstract}
  The amoeboid organism, the plasmodium of
  {\it Physarum polycephalum}, behaves on the basis of spatio-temporal
  pattern formation by local contraction-oscillators. This biological
  system can be regarded as a reaction-diffusion system which has
  spatial interaction by active flow of protoplasmic sol in the
  cell. Paying attention to the physiological evidence that
  the flow is determined by contraction pattern in the plasmodium, a
  reaction-diffusion system having self-determined flow arises. Such
  a coupling of reaction-diffusion-advection is a characteristic of
  the biological system, and is expected to relate with control
  mechanism of amoeboid behaviours. Hence, we have studied effects of
  the self-determined flow on pattern formation of simple
  reaction-diffusion systems. By weakly nonlinear analysis near a
  trivial solution, the envelope dynamics follows the complex
  Ginzburg-Landau type equation just after bifurcation occurs at
  finite wave number. The flow term affects the nonlinear term of
  the equation through the critical wave number squared. Contrary to
  this, wave number isn't explicitly effective with lack of flow or
  constant flow. Thus, spatial size of pattern is especially important
  for regulating pattern formation in the plasmodium. On the other
  hand, the flow term is negligible in the vicinity of bifurcation at
  infinitely small wave number, and therefore the pattern formation by 
  simple reaction-diffusion will also hold. A physiological role of
  pattern formation as above is discussed.
\end{abstract}



\section{Introduction}
\label{sec: Introduction}

The plasmodium of {\it Physarum polycephalum\/} is a large amoeboid
cell, showing contraction-relaxation cycles everywhere within an
organism. These local contraction motivates intracellular transport of
endoplasmic sol \cite{Kamiya59}.
Therefore, the plasmodium can crawl when the local flow is
appropriately arranged all through the organism. Some types of
spatio-temporal pattern of the contraction have been observed after
stimulation, and discussed in relation to development of its amoeboid
behaviours \cite{Matsumoto86,Ueda93,Miyake96,Nakagaki96}.
Mechanism of the pattern formation is closely related to a mechanism
of cellular behaviours in the plasmodium. Then we expect that a
certain origin of biological fineness for controlling a system is in
characteristic mechanism of the biological pattern formation. This
paper is concerned with such character in a biological system.

Let us consider theoretical framework of the pattern formation in the 
plasmodium. Chemical oscillation is a clock of the rhythmic
contraction. Possible candidates for the chemicals are, for example,
Ca${}^{2+}$ and/or ATP
\cite{Ueda93,Yoshimoto81,Ogihara82}.
Since the contraction apparatus is located at outer layer of the
plasmodium, called ectoplasm
\cite{Ueda78,NaibMajani82,Ishigami86},
it is proper to focus on chemical oscillations and diffusion in the
ectoplasm. As above, the plasmodial contraction system can be regarded
as diffusively coupled oscillators or a reaction-diffusion system in
the ectoplasm. Actually, some experimental results can be explained
using this framework
\cite{Takahashi97,Takamatsu97}.

But another effect of spatial interaction must be considered. That is
the protoplasmic streaming in inner part of the plasmodium, called
endoplasm, because the contraction pattern is modified by inhibition
of the streaming \cite{Nakagaki96,Yoshimoto78,Miyake91}.
Miyake {\it et al.\/} \cite{Miyake91} proposed a
model of the information processing system with two levels of
subsystems corresponding to the endoplasmic oscillators with
long-range interaction and the ectoplasmic ones with short-range
interaction. From more physical viewpoints, some models based on
hydrodynamics and chemical kinetics, have been presented
\cite{Oster84,Teplov,Smith95}.
However, we still have an issue that the streaming can be sometimes
negligible, as mentioned in the former paragraph. To solve the
pretence of this discrepancy, it is inevitable to study pattern
formation including the streaming effect. For that purpose, we firstly 
suppose a physical model of reaction-diffusion-advection equations.

The oscillating chemicals are exchanged between endo- and ectoplasm,
and flow out/in via the streaming
\cite{Grebecki78,Baranowski82}.
The streaming is motivated by the gradient of contraction force
\cite{Kamiya72}.
Hence, the contraction pattern determines the streaming. That is to
say, the protoplasmic streaming is a factor of the contraction
pattern, and also inversely motivated by the pattern.
Then the reaction-diffusion system having self-determined flow arises.
The self-determined flow is just a characteristic of the biological
system in the plasmodium. We analyze pattern formation of such a
system by weakly nonlinear analysis, particularly noticing effects of
the advection term on simple reaction-diffusion systems. The weakly
nonlinear analysis is based on singular perturbation expansion, and
therefore has an advantage that model equations are free from details.
Because, no matter how complex or uncertain the functions are in our
model equations, expanded lower order terms are merely considered in
the analysis. This advantage is valuable especially in analysis of
biological model, which often includes uncertain assumptions in
detail.

As the results of the analysis, a possible solution of the pretence
discrepancy described earlier is found, and furthermore, one apparent
character of the plasmodial pattern formation is shown. A possible
role of the plasmodial pattern formation is discussed at a
physiological point of view.


\section{Basic equations}
\label{sec: Basic equations}


The plasmodium of {\it Physarum\/} has a cytoplasmic cortex
(ectoplasmic gel) filled with endoplasmic sol. The ectoplasm makes
periodic contraction and relaxation, and it causes intracellular
streaming of the endoplasm. A sheet of cytoplasm become thick when the
endoplasm flowing into it. Metabolic substances regulate contractile 
cycles, diffuse in the cytoplasm, and are transferred by the
endoplasmic flow. Standing on these points, we present a dynamics of
rhythmic pattern formation in the {\it Physarum\/} plasmodium as the
following equations for metabolic elements in the ectoplasmic gel,
${\bf u}_{\rm gel}$ and endoplasmic sol, ${\bf u}_{\rm sol}$:
\begin{eqnarray}
  &&\frac{\partial h}{\partial t}
  + \vec{\nabla} \cdot (h \vec{v})
  = 0,
  \nonumber\\
  &&\frac{\partial {\bf u}_{\rm gel}}{\partial t}
  = {\bf F}_{\rm gel}(h, {\bf u})
  + \vec{\nabla} \cdot (D_{\rm gel} \vec{\nabla} {\bf u}_{\rm gel}),
  \label{eqn: RDA}\\
  &&\frac{\partial {\bf u}_{\rm sol}}{\partial t}
  + \vec{v} \cdot \vec{\nabla} {\bf u}_{\rm sol}
  = {\bf F}_{\rm sol}(h, {\bf u})
  + \frac{1}{h}
  \vec{\nabla} \cdot (D_{\rm sol} \vec{\nabla} h {\bf u}_{\rm sol}).
  \nonumber
\end{eqnarray}
Where $h$ denotes the thickness of the endoplasmic sol, and
$\vec{v}$ is the averaged velocity of the endoplasmic flow.
${\bf u}$ is $N$ metabolic species in gel and sol,
$({\bf u}_{\rm gel}, {\bf u}_{\rm sol}) = (u^1, \ldots, u^N)$.
${\bf F}_{\rm gel}$ and ${\bf F}_{\rm sol}$ represent reaction kinetics
among metabolic elements and exchanges of them between the ectoplasm
and endoplasm. $D_{\rm gel}$ and $D_{\rm sol}$ are diagonal matrices
of diffusion constants of metabolic elements.
We note that Eq.~(\ref{eqn: RDA}) becomes the closed system if the
dynamics of $\vec{v}$, the equation of motion of endoplasmic flow, are
given. Let us consider the small Reynolds number flow in a contractile 
vein and assume the intracellular pressure is determined by the
concentration of the chemicals:
$\vec{v} = - q(h) \vec{\nabla} P({\bf u})$.
This means the stationary flow approximation that deformation of the
ectoplasm is very slow compared with variation of the endoplasmic flow.
It is hence not considered that the ectoplasmic cortex has
viscoelastic features.


In Eq.~(\ref{eqn: RDA}), the sol-gel conversion is ignored, and thus
the mass conservation of endoplasmic sol is satisfied. This implies
the limitation of the model, such as the cell motility, the formation
and reconnection of a network of protoplasmic strands.
In the following, we assume that the thickness of the endoplasm is
almost constant all over the plasmodium, and that the diffusion
constants of the metabolic elements are homogeneous in the
plasmodium. The intracellular pressure $P$ is expanded around the
homogeneous static state ${\bf u} = {\bf u}_s$ as
\begin{displaymath}
  P({\bf u}) = P({\bf u}_s)
  + \sum_j (u^j - u^j_s) \frac{\partial P}{\partial u^j}({\bf u}_s)
  + \mbox{higher order terms}.
\end{displaymath}
Hereafter, we ignore the higher order terms which have no effects on
results deduced by the weakly nonlinear analysis. Under these
assumptions, we rewrite Eq.~(\ref{eqn: RDA}) in the form of the
reaction-diffusion-advection equations:
\begin{equation}
  \frac{\partial {\bf u}}{\partial t}
  + M \vec{\nabla} {\bf u} \cdot \vec{\nabla} {\bf u}
  = {\bf F}({\bf u}; \mu) + D \vec{\nabla}^2 {\bf u},
  \label{eqn: rda}
\end{equation}
where ${\bf F}$ is reaction kinetics, $M$ represents a tensor of
advection coefficients induced by endoplasmic flow, and $D$ is a
diagonal matrix of diffusion constants.


We assume that the system~(\ref{eqn: rda}) has a trivial homogeneous
steady solution ${\bf u} = {\bf o}$, and it is stable when the
bifucation parameter is $\mu < \mu_c$. Over the bifurcation
point, $\mu > \mu_c$, the trivial solution is unstable and local
reaction kinetics becomes oscillatory. Reaction kinetics 
$F({\bf u}; \mu)$ can be expanded around the trivial solution as
\begin{eqnarray*}
  {\bf F}({\bf u}; \mu)
  &=& L {\bf u} + N_2 {\bf u} {\bf u} + N_3 {\bf u} {\bf u} {\bf u} + \cdots,
  \\
  L &=& \Bigl( \frac{\partial F^i}{\partial u^j} \Bigr),\;\;
  N_2 = \Bigl(
  \frac{\partial^2 F^i}{\partial u^j \partial u^k}
  \Bigr),\;\;
  N_3 = \Bigl(
  \frac{\partial^3 F^i}{\partial u^j \partial u^k \partial u^l}
  \Bigr),\ldots
  \;\; \mbox{at} \;\; {\bf u} = {\bf o}.
\end{eqnarray*}


Linearizing the system~(\ref{eqn: rda}) near the homogeneous steady
state, we obtain equations for spatial Fourier components,
\begin{displaymath}
  \dot{{\bf W}}(t; k) = (L - k^2 D) {\bf W}(t; k),\quad
  {\bf u}(t, x) = \int {\bf W}(t; k) e^{i k x} d k.
\end{displaymath}
The stability of ${\bf W} = {\bf o}$ is given by the eigenvalue problem
for the eigenvalue $\lambda$ and eigenvector ${\bf U}$,
\begin{equation}
  (L - k^2 D) {\bf U} = \lambda {\bf U}.
  \label{eqn: L}
\end{equation}
We obtain the eigenvalue 
$\lambda = ({\bf U}^*, (L - k^2 D) {\bf U})/({\bf U}^*, {\bf U})$ 
from Eq.~(\ref{eqn: L}).
On the bifucation point, $\mbox{Re} \lambda = 0$ and
$\partial (\mbox{Re} \lambda) / \partial k = 0$ are satisfied at some
$k = k_c$ for the maximal eigenvalue(s). In the vicinity of the
bifurcation point $(\mu, k) = (\mu_c, k_c)$, the eigenvalue is
\begin{equation}
  \lambda
  = \lambda_c
  + \frac{\partial \lambda}{\partial \mu}
  \Big|_c (\mu - \mu_c)
  + \frac{\partial \lambda}{\partial k}
  \Big|_c (k - k_c)
  + \frac{1}{2!} \frac{\partial^2 \lambda}{\partial k^2}
  \Big|_c (k - k_c)^2
  + \cdots,
  \label{eqn: dispersion}
\end{equation}
where the subscript $c$ denotes the bifurcation point, and
expansion coefficients are given as
\begin{displaymath}
  \lambda_c = \pm i \omega_c,\quad
  \frac{\partial \lambda}{\partial k}\Big|_c = \pm i c_g,\quad
  \omega_c, c_g \geq 0.
\end{displaymath}
Hence the bifurcation problem has four
generic cases: (i) $k_c = 0$ and  $\omega_c = 0$;
(ii) $k_c = 0$ and  $\omega_c \neq 0$ (Hopf bifurcation);
(iii) $k_c \neq 0$ and  $\omega_c = 0$ (Turing bifurcation);
(iv) $k_c \neq 0$ and  $\omega_c \neq 0$ (travelling-wave type).
According to the weakly nonlinear analysis of Eq.~(\ref{eqn: rda}),
the advection term is expected to modulate its bifurcation behaviour
in the case of (iii) and (iv).


\section{Envelope equation}
\label{sec: Envelope equation}


Let us consider the envelope equation just after the
bifurcation of travelling-wave type occurs, on a basis of weakly
nonlinear analysis \cite{WNA}.
We denotes the bifurcation parameter by
$(\mu - \mu_c) \sim \epsilon^2$, and suppose that
$u^j \sim O(\epsilon)$ ($j = 1, \ldots, N$) in the vicinity of the
bifurcation point.
In the following analysis, the envelope equation is derived for
one spatial dimension system of Eq.~(\ref{eqn: rda}) with the single
travelling wave. We introduce perturbation expansions and multiple
scales,
\begin{eqnarray}
  &&{\bf u} \sim \epsilon {\bf u}_1 + \epsilon^2 {\bf u}_2
  + \epsilon^3 {\bf u}_3 + \cdots, \quad
  L \sim L_0 + \epsilon^2 L_2 + \cdots,
  \nonumber\\
  &&X = x - c_p t,\quad
  \xi = \epsilon (x - c_g t),\quad
  \tau = \epsilon^2 t,
  \label{eqn: expansions}
\end{eqnarray}
where $c_p = \omega_c / k_c$ is the phase velocity and $c_g$ is the
group velocity.
Substitution of Eq.~(\ref{eqn: expansions}) into~(\ref{eqn: rda})
yields perturbation equations for each order in $\epsilon$:
\begin{equation}
  O(\epsilon^m)\qquad
  \Bigl( L_0 + D \frac{\partial^2}{\partial X^2}
  + c_p \frac{\partial}{\partial X} \Bigr)
  {\bf u}_m = {\bf b}_m,\quad
  m = 1, 2, 3, \ldots
  \label{eqn: mth}
\end{equation}
where ${\bf b}_m$ denotes the inhomogeneous term of the $m$-th order
equation.

For the first order equation in Eq.~(\ref{eqn: mth}), the
inhomogeneous term is ${\bf b}_1 = {\bf o}$. Then we have a solution
\begin{displaymath}
  {\bf u}_1 = W(\xi, \tau) e^{i k_c X} {\bf U} + \mbox{c.c.},
\end{displaymath}
where {c.c.} means complex conjugate. ${\bf U}$ is an eigenvector of
an eigenvalue $\lambda = - i \omega_c$ for the eigenvalue
problem~(\ref{eqn: L}) on the bifurcation point.

For the second order equation, that is $m = 2$ in
Eq.~(\ref{eqn: mth}), we expand the solution and inhomogeneous term by
the phase $\phi = k_c x - \omega_c t$,
\begin{displaymath}
  {\bf u}_2 = \sum_l {\bf u}_2^{(l)} e^{i l \phi}, \quad
  {\bf b}_2 = \sum_l {\bf b}_2^{(l)} e^{i l \phi}.
\end{displaymath}
Then the solvability conditions for ${\bf u}_2$ are
\begin{eqnarray}
  ({\bf U}^*, {\bf b}_2^{(+1)})
  &=&
  (\bar{\bf U}^*, {\bf b}_2^{(-1)})
  = 0,
  \label{eqn: sov of 2nd}\\
  {\bf b}_2^{(+1)}
  &=&
  - \frac{\partial W}{\partial \xi} (c_g + 2 i k_c D) {\bf U},
  \quad
  {\bf b}_2^{(-1)}
  = \bar{\bf b}_2^{(+1)}.
  \nonumber
\end{eqnarray}
These conditions are obviously satisfied since
\begin{displaymath}
  c_g = i \frac{\partial \lambda}{\partial k}\Big|_c
  = -2 i k_c \frac{({\bf U}^*, D {\bf U})}{({\bf U}^*, {\bf U})}.
\end{displaymath}
Thus we must advance our calculation to
the third order to obtain the envelope equation.

For the third order equation in $\epsilon$, we expand ${\bf u}_3$ and
${\bf b}_3$ as
\begin{displaymath}
  {\bf u}_3 = \sum_l {\bf u}_3^{(l)} e^{i l \phi}, \quad
  {\bf b}_3 = \sum_l {\bf b}_3^{(l)} e^{i l \phi},
\end{displaymath}
then the solvability conditions for ${\bf u}_3$,
\begin{eqnarray}
  ({\bf U}^*, {\bf b}_3^{(+1)})
  &=& (\bar{\bf U}^*, {\bf b}_3^{(-1)})
  = 0,
  \label{eqn: sov of 3rd}\\
  {\bf b}_3^{(+1)}
  &=& \frac{\partial W}{\partial \tau} {\bf U}
  - c_g \frac{\partial {\bf u}_2^{(1)}}{\partial \xi}
  + 2 k_c^2 \bar{W} M ({\bf u}_2^{(2)} \bar{\bf U}
  + \bar{\bf U} {\bf u}_2^{(2)})
  - W L_2 {\bf U}
  \nonumber\\
  && - W N_2 ({\bf u}_2^{(0)} {\bf U} + {\bf U} {\bf u}_2^{(0)})
  - \bar{W} N_2 ({\bf u}_2^{(2)} \bar{\bf U}
  + \bar{\bf U} {\bf u}_2^{(2)})
  \nonumber\\
  && - |W|^2 W N_3 ({\bf U} {\bf U} \bar{\bf U}
  + {\bf U} \bar{\bf U} {\bf U} + \bar{\bf U} {\bf U} {\bf U})
  - 2 i k_c D \frac{\partial {\bf u}_2^{(1)}}{\partial \xi}
  - \frac{\partial^2 W}{\partial \xi^2} D {\bf U},
  \nonumber\\
  {\bf b}_3^{(-1)}
  &=& \bar{\bf b}_3^{(+1)}.
  \nonumber
\end{eqnarray}
Thus we obtain the complex Ginzburg-Landau (CGL) equation:
\begin{eqnarray}
  \frac{\partial W}{\partial t}
  &=& c_1 W - c_2 |W|^2 W + c_3 \frac{\partial^2 W}{\partial \xi^2},
  \quad c_j = ({\bf U}^*, {\bf V}_j)/({\bf U}^*, {\bf U}),
  \label{eqn: CGL}\\
  {\bf V}_1
  &=& L_2 {\bf U},
  \nonumber\\
  {\bf V}_2
  &=& - 2 k_c^2 M
  \Bigl\{ \bar{\bf U},
  \bigl(L_0^{(2, 2)}\bigr)^{-1} \bigl(N_2^{(+1)} {\bf U} {\bf U}\bigr)
  \Bigr\}
  + N_2
  \Bigl\{ {\bf U},
  L_0^{-1} \bigl(N_2^{(-1)} \{{\bf U}, \bar{\bf U}\}\bigr)
  \Bigr\}
  \nonumber\\
  && + N_2
  \Bigl\{ \bar{\bf U},
  \bigl(L_0^{(2, 2)}\bigr)^{-1} \bigl(N_2^{(+1)} {\bf U} {\bf U}\bigr)
  \Bigr\}
  - N_3
  \Bigl( {\bf U} {\bf U} \bar{\bf U}
  + {\bf U} \bar{\bf U} {\bf U}
  + \bar{\bf U} {\bf U} {\bf U} \Bigr),
  \nonumber\\
  {\bf V}_3
  &=&
  - (c_g + 2 i k_c D) (L_0^{(1, 1)})^{-1} (c_g + 2 i k_c D) {\bf U},
  \nonumber
\end{eqnarray}
where $L_0^{(l, m)} = L_0 - (l k_c)^2 D + i m \omega_c$,
$N_2^{(l)} = N_2 + l k_c^2 M$, and $\{ X, Y \} = X Y + Y X$.
By means of the dispersion relation~(\ref{eqn: dispersion}), 
$c_1 = (\partial \lambda / \partial \mu)_c$ and
$c_3 = (1/2) (\partial^2 \lambda / \partial k^2)_c$ are generically
complex constants.

The CGL equation~(\ref{eqn: CGL}) describes the small amplitude
dynamics of the system near the bifurcation point, that is, $W$ slowly
and slightly modulates travelling wave with the wave number $k_c$ and
frequency $\omega_c$. The coefficient of the nonlinear term, $c_2$,
depends on the advection term in Eq.~(\ref{eqn: rda}). Since the
advection term has the form of the gradient of metabolic species,
$c_2$ is including the term of $k_c^2$. It is known that $c_2$
determines the amplitude of plane wave solutions and the nonlinear
dispersion relation. The signature of $\mbox{Re}(c_2)$ decides the
type of bifurcation at $\mbox{Re}(c_1) = 0$: a supercritical
bifurcation occurs for $\mbox{Re}(c_2) > 0$, while subcritical one for
$\mbox{Re}(c_2) < 0$. In the latter case, we will need higher order
terms as $|W|^4 W$ to the CGL equation.

The CGL equation has various type of solutions with relation to the
coefficients \cite{CGL},
but we do not mention each of them. Because we have
derived the CGL equation from Eq.~(\ref{eqn: rda}) which has
no concrete form of reaction and advection terms, it is
impossible to give the coefficients of the CGL equation explicitly.
In the next section, we discuss the advection effect implied by the
CGL equation~(\ref{eqn: CGL}) and compare our results with ones of
reaction-diffusion models without the endoplasmic flow.

We comment on the envelope equation of spatial two dimensions.
Add a scaling of $y$-axis, $\eta = \epsilon y$, to
Eq.~(\ref{eqn: expansions}) and continue the similar calculation
above, we obtain the CGL-type equation without rotational symmetry
(see appendix).


\section{Conclusions}
\label{sec: conclusions}


We have presented a reaction-diffusion-advection model of
the {\it Physarum\/} plasmodium which makes rhythmic contraction. The
local contraction determines intracellular transport of endoplasmic
sol, and this motivates the crawling behaviour of the plasmodium.
By means of the weakly nonlinear analysis of the model, we obtain the
following results:
(1) an advection effect arises to the slow dynamics when the
travelling-wave type bifurcation occurs;
(2) the envelope equation has the form of the CGL equation;
(3) the advection terms affect the nonlinear term of the CGL equation
through $k_c^2$.
In weakly nonlinear region of a reaction-diffusion-advection model,
(2) implies that we can observe similar dynamical behaviour to
conventional reaction-diffusion systems without flow. Thus it is
expected that some experimental results of the rhythmic contraction
in the plasmodium have been illustrated by such simple
reaction-diffusion systems. The last result (3), however, shows
possibility that the nonlinear effect near the bifurcation point
stems from not only reaction kinetics but also the self-determined
flow. In the remainder of this section, we attempt to discuss the
effect of the self-determined flow on physiological behaviour.


In general, advection can play an effective role for pattern
formation, as the flow of matter often causes instabilities of 
hydrodynamical systems \cite{Ei90,Perumpanani95}.
For example, Rovinsky and Menzinger \cite{Rovinsky} have shown
that a differential flow of chemical species induces instabilities of
the homogeneous steady state, and leads to travelling wave pattern
nevertheless without diffusions. In this case, advection terms have a
crucial effect on linear dispersion relation of the wave propagation.
Contrary to this, the advection studied in the present paper, the
self-determined flow, has no effect on the linear stability, but
modulates nonlinear dispersion relation.


In weakly nonlinear region, the self-determined advection causes the
strong dependency of pattern formation on the critical wave number,
$k_c$. The affected nonlinear term of the CGL equation is known to be 
closely related with bifurcation and size of amplitude. Because phase
difference of the metabolic oscillation between neighbours tends to
become larger as $k_c$ is larger, $k_c$ is regarded as an indicator
that an oscillator is in step with its neighbouring oscillators.
This point of view implies that phase difference in local plays an
important role for pattern formation such as amoeboid behaviours in
the {\it Physarum\/} plasmodium.

Real patterns observed by measuring thickness oscillation in the
plasmodium show that the wave number is always about one or two in
free locomotion under culture conditions. However, larger wave number
more than ten appears transiently in relation to changes in behaviours 
induced by an appropriate stimulation. This phenomenon may relate to
the $k_c$-dependency. Further analysis of the $k_c$-effect would be
given if the model equations are concretely specified.


\section*{Acknowledgments}

We thanks K. Imai for helpful comment through discussion with H. Y.
This study was supported by The Sumitomo Foundation (\#970628), and
The Institute of Physical and Chemical Research (RIKEN) as a special
post-doctoral researcher for basic science (assigned to T. N.).

\appendix

\section*{Derivation of the envelope equation in two dimensions}
\label{sec: EE}


Before studying the envelope equation in two dimensions, we comment
on counter propagating waves.
For the bifurcation of travelling-wave type, counter propagating waves
are possible in one spatial dimension, although we derived the
envelope equation (CGL equation) for the single travelling wave in
section~\ref{sec: Envelope equation}.
The envelope dynamics of the counter propagating waves is the coupled
equations of envelopes $W_+$ and $W_-$ of a linearlized solution
\begin{displaymath}
  W_+ e^{i (k_c x + \omega_c t)} {\bf U}_+
  + W_- e^{i (k_c x - \omega_c t)} {\bf U}_-
  + \mbox{c.c.},
\end{displaymath}
where $(L_0 - k_c^2 D) {\bf U}_\pm = \pm i \omega_c {\bf U}_\pm$. The
time evolution of $W_\pm$ is governed by the dynamics like CGL
equation, but is modulated by the interaction terms
$|W_\mp|^2 W_\pm$.


In the case of two spatial dimensions, multimode travelling waves
satisfying $|{\bf k}_j| = k_c$,
\begin{displaymath}
  \sum_j
  \Bigl(
  W_{j+} {\bf U}_+
  \exp\bigl[ i({\bf k}_j \cdot {\bf x} + \omega_c t) \bigr]
  + W_{j-} {\bf U}_-
  \exp\bigl[ i({\bf k}_j \cdot {\bf x} - \omega_c t) \bigr]
  \Bigr) + \mbox{c.c.}
\end{displaymath}
are possible because modes corresponding to an anulus of wave vectors,
$|{\bf k}| = k_c$, are neutrally stable on the bifurcation point.
Although we can obtain the multimode envelope equations, 
it is not clear that the dynamics of them has structural stability.
Such a problem has a relation to pattern selection, and we need more
precise analysis on the bifurcation with symmetry. Hereafter we
discuss the envelope equation for a single travelling wave of
Eq.~(\ref{eqn: rda}) in two spatial dimensions. We assume that a set
of mode near the single plane wave travelling along $x$-axis,
$\exp[i(k_c x - \omega_c t)]$, is dominant across the bifurcation
point.


Introducing perturbation expansions and multiple scales,
\begin{eqnarray}
  &&{\bf u} \sim \epsilon {\bf u}_1 + \epsilon^2 {\bf u}_2
  + \epsilon^3 {\bf u}_3 + \cdots, \quad
  L \sim L_0 + \epsilon^2 L_2 + \cdots,
  \nonumber\\
  &&X = x - c_p t,\quad
  \xi = \epsilon (x - c_g t),\quad
  \eta = \epsilon y,\quad
  \tau = \epsilon^2 t,
  \label{eqn: exp}
\end{eqnarray}
and substituting Eq.~(\ref{eqn: exp}) into Eq.~(\ref{eqn: rda}),
we obtain perturbation equations,
\begin{equation}
  O(\epsilon^m)\qquad
  \Bigl( L_0 + D \frac{\partial^2}{\partial X^2}
  + c_p \frac{\partial}{\partial X} \Bigr)
  {\bf u}_m = {\bf b}_m,\quad
  m = 1, 2, 3, \ldots
  \label{eqn: m}
\end{equation}
The inhomogeneous term of the first order equation is vanishing,
${\bf b}_1 = {\bf o}$. For the second and third order equations, the
inhomogeneous terms are
\begin{eqnarray*}
  {\bf b}_2 &=&
  - c_g \frac{\partial {\bf u}_1}{\partial \xi}
  + M \frac{\partial {\bf u}_1}{\partial X}
  \frac{\partial {\bf u}_1}{\partial X}
  - N_2 {\bf u}_1 {\bf u}_1
  - 2 D \frac{\partial^2 {\bf u}_1}{\partial X \partial \xi},
  \\
  {\bf b}_3 &=&
  \frac{\partial {\bf u}_1}{\partial \tau}
  - c_g \frac{\partial {\bf u}_2}{\partial \xi}
  + M \Bigl(
  \frac{\partial {\bf u}_2}{\partial X}
  \frac{\partial {\bf u}_1}{\partial X} +
  \frac{\partial {\bf u}_1}{\partial X}
  \frac{\partial {\bf u}_2}{\partial X} +
  \frac{\partial {\bf u}_1}{\partial \xi}
  \frac{\partial {\bf u}_1}{\partial X} +
  \frac{\partial {\bf u}_1}{\partial X}
  \frac{\partial {\bf u}_1}{\partial \xi}
  \Bigr)
  \\
  && - L_2 {\bf u}_1
  - N_2 ({\bf u}_2 {\bf u}_1 + {\bf u}_1 {\bf u}_2)
  - N_3 {\bf u}_1 {\bf u}_1 {\bf u}_1
  - D \Bigl(
  2 \frac{\partial^2 {\bf u}_2}{\partial X \partial \xi}
  + \frac{\partial^2 {\bf u}_1}{\partial \xi^2}
  + \frac{\partial^2 {\bf u}_1}{\partial \eta^2}
  \Bigr).
\end{eqnarray*}
We write a solution of the first order equations with the slowly
varying envelope $W(\xi, \eta, \tau)$,
\begin{displaymath}
  {\bf u}_1 = W e^{i k_c X} {\bf U} + \mbox{c.c.}
\end{displaymath}
Since the solvability conditions for the second order equation
are satisfied by the definition of $c_g$,
we advance our calculation to the third order:
\begin{eqnarray}
  ({\bf U}^*, {\bf b}_3^{(+1)})
  &=& (\bar{\bf U}^*, {\bf b}_3^{(-1)})
  = 0,
  \label{eqn: sov3}\\
  {\bf b}_3^{(+1)}
  &=& \frac{\partial W}{\partial \tau} {\bf U}
  - c_g \frac{\partial {\bf u}_2^{(1)}}{\partial \xi}
  + 2 k_c^2 \bar{W} M ({\bf u}_2^{(2)} \bar{\bf U}
  + \bar{\bf U} {\bf u}_2^{(2)})
  - W L_2 {\bf U}
  \nonumber\\
  && - W N_2 ({\bf u}_2^{(0)} {\bf U} + {\bf U} {\bf u}_2^{(0)})
  - \bar{W} N_2 ({\bf u}_2^{(2)} \bar{\bf U}
  + \bar{\bf U} {\bf u}_2^{(2)})
  \nonumber\\
  && - |W|^2 W N_3 ({\bf U} {\bf U} \bar{\bf U}
  + {\bf U} \bar{\bf U} {\bf U} + \bar{\bf U} {\bf U} {\bf U})
  - 2 i k_c D \frac{\partial {\bf u}_2^{(1)}}{\partial \xi}
  - \Bigl(
  \frac{\partial^2 W}{\partial \xi^2}
  + \frac{\partial^2 W}{\partial \eta^2}
  \Bigr) D {\bf U},
  \nonumber\\
  {\bf b}_3^{(-1)}
  &=& \bar{\bf b}_3^{(+1)}.
  \nonumber
\end{eqnarray}
From the solvability conditions (\ref{eqn: sov3}), we get the envelope
dynamics,
\begin{equation}
  \frac{\partial W}{\partial \tau}
  = c_1 W - c_2 |W|^2 W
  + (c_3 + c_4) \frac{\partial^2 W}{\partial \xi^2}
  + c_4 \frac{\partial^2 W}{\partial \eta^2},
  \label{eqn: 2dCGL}
\end{equation}
where coefficients $c_1$, $c_2$, $c_3$ are the same in
Eq.~(\ref{eqn: CGL}) and $c_4 = - i c_g / 2 k_c$. The coefficient
$c_4$ implies dipersion waves which are observed in the nonlinear
Schr\"odinger equation.



\end{document}